\documentclass[conference]{IEEEtran}
\usepackage[utf8]{inputenc}
\usepackage{cite}
\usepackage{amsmath,amssymb,amsfonts}
\usepackage{graphicx}
\usepackage{subfigure}
\usepackage{textcomp}
\usepackage{forest}
\usepackage{xcolor}
\usepackage{algorithm,algorithmic}
\usepackage{bm}
\usepackage{booktabs}
\usepackage{float}
\usepackage{cleveref}
\usepackage{pgf, tikz}
\usepackage{booktabs}
\usepackage{float}
\usetikzlibrary{arrows, automata}
\usepackage{etoolbox}
\usetikzlibrary{matrix,positioning}
\usepackage{algorithmic}
\usepackage{graphicx}
\usepackage{graphics}
\usepackage{textcomp}
\usepackage{subfigure}
\usepackage{tabularx}
\usepackage{times}
\usepackage{multirow}
\usepackage{bm}
\usepackage{latexsym}
\usepackage{booktabs}
\usepackage{threeparttable}
\usepackage{epsf}
\usepackage{algorithm}
\usepackage{color,colortbl}
\usepackage{tabularx}
\usepackage{makecell}
\usepackage{pgfplots}
\usepackage{url}

\usepackage{verbatim}

\title{Fast Distribution Grid Topology Estimation\\ via Subset Sum}

\author{\IEEEauthorblockN{Yueyao Xu}
    \IEEEauthorblockA{Hong Kong Polytechnic University \\
    Hong Kong SAR
    \\michelle.xu@connect.polyu.hk}
    \and
    \IEEEauthorblockN{
Yize Chen}
    \IEEEauthorblockA{University of Alberta\\
    Edmonton, Canada
    \\yize.chen@ualberta.ca}
}

\begin{document}

\maketitle

\begin{abstract}
Faced with increasing penetration of distributed energy resources and fast development of distribution grid energy management, topology identification of distribution grid becomes an important and fundamental task. As the underlying grid topology is usually unknown or incomplete to the utilities, it is becoming a fundamental task to efficiently identify the distribution grid network topology using limited measurements. A fast and accurate topology identification can help achieving the tasks of load monitoring, operation and control of power distribution system as well as outage detection. In this paper, we propose a novel and ultra-fast topology identification method. By adapting the subset sum method with a hierarchical structure, the overall grid topology can be inferred from fewer samples of smart meter power measurements. Such techniques can be applied in real time under the scenarios with fast topology change, and the proposed hierarchical algorithm is also robust against measurement noises.

\end{abstract}

\section{Introduction}
With the development of technology and the addition of emerging distributed energy resources and plug-and-play appliances, the distribution grid is witnessing an increasing level of complexity. In transmission networks, control and monitoring of power grids are based on known topology and line parameters. However, the topology of distribution grid is time varying and often unknown to utility operations. And with the sharp increase in the number of low-voltage distribution users and distributed energy resources (DERs), while traditional regulatory methods require a large number of personnel, it is increasingly difficult to monitor the topology and energy usage in low-voltage distribution networks \cite{william2002distribution, fan2009evolution}.

Accurate network topology information plays an important role in detection, location, state estimation and optimization of power systems~\cite{deka2017structure}. The topology of a low-voltage distribution network gives the connectivity among various kind of assets such as feeders, distribution transformers, distributors and consumers \cite{lueken2012distribution}. Efficient topology identification techniques can also reduce technical losses from grid changes. Because the power grid changes during reconfiguration, repair, and maintenance, this phenomenon may cause network topology information not to be accurately available at fast timescales \cite{chembe2009reduction}. Therefore, it is difficult for network operators to realize these changes in topology and find problems of the power grid in time. At the same time, it is nontrivial for distribution companies to keep accurate records of the topology. 

With recent advances and deployment of Advanced Metering Infrastructure
(AMI), intelligent electric meters can transmit readings at higher frequencies to a central data center in real time \cite{jayadev2016novel}. Meanwhile, the observability of the power grid can be enhanced as utility companies are installing smart meters at important nodes of the network, including feeders, transformers, and demand nodes. Moreover, though the design of distribution networks may appear to be loopy or meshed, the vast majority of distribution grids are operated as ``radial” networks with a set of non-overlapping trees. Switches in the network are used to achieve one radial configuration out of many possibilities. And in such cases, the distribution grid topology recovery can be treated as identifying all the trees' connections given a specified switch configuration. These characteristics combined together make it possible for us to design fast topology identification schemes to improve the topology recovery accuracy and efficiency.

In order to tackle such challenge to distribution grid, many
attempts have been made. Statistical methods such as approaches based on covariance matrix and Bayesian analysis are proposed to model and infer the network topology based on nodal conditional relationship~\cite{deka2017structure, mabrouk2022distribution}. In \cite{li2013blind}, the sparsity of grid parameters are considered, and an optimization based to approach is developed for reconstructing the graph Laplacian. \cite{ardakanian2019identification} develops adaptive lasso penalty to estimate large elements of the admittance matrix, and a joint problem with state estimation has been also considered~\cite{karimi2021joint}. Furthermore, \cite{cavraro2018graph} considers active probing approach and discusses topology recoverability for distribution feeder.  Yet some necessary information, such as locations of switches or admittance matrices, are not available in practical distribution networks. Another line of topology identification methods integrate Principal component analysis~(PCA) and its graph theory interpretation. Based on the principle of power balance, the network topology is identified from the time series of power measurements \cite{pappu2017identifying}.
However, PCA method does not realize the recognition of multi-layer complex power grid topology, and dimension reduction still takes a long time for a large number of samples. At the same time, PCA algorithm may combine nodes with similar statistical features, which could make the recovered topology inaccurate.


In this paper,  we propose a straightforward but highly effective approach for power grid topology identification. Our  method is based on the subset sum algorithm and graph theory. Based on the observation of radial network structure and summation relations of nodal power measurements, we adapt the classical combinatorial algorithm to better accommodate tree structure in distribution grid. The proposed method can be well generalized to larger systems, and makes minimal assumption on the known information of the underlying grid. With the utilization of nodal power measurements, high accuracy result of identifying the power grid topology  can be achieved with minimal sample requirements, and the result is compared with the algorithm based on PCA. 
The rest of the paper is organized as follows. Section II revisits some necessary preliminaries and The mathematical formulation of the problem. The proposed solution and algorithms are presented in Section III. Finally, the simulation results with time complexity analysis and conclusions are provided in Sections IV and V, respectively.

\section{Distribution Grid Topology Recovery}
\subsection{Problem Setup}
We are interested in recovering radial distribution grid topology using measurement data. For a distribution grid $\mathcal{G} = (\mathcal{N}, \mathcal{E})$, where $\mathcal{N}$ is a set of $n$ nodes; $\mathcal{E}$ is a set of distribution lines, which are unordered pairs of vertices. We assume in the underlying distribution grid, it does not allow multiple edges between same pair of vertices. For each node $j$, we assume there is a complex power injection $s_j=p_j+i q_j$ with voltage magnitude $v_j=|V_j|^2$. For each power line $jk$ with active power flow $P_{jk}$ and reactive power flow $Q_{jk}$, denote the complex impedance as $z_{jk}=r_{jk}+i x_{jk}$.  In this paper, the linearized DistFlow model is adopted to map the power flow and the electrical parameters of the grid~\cite{baran1989optimal}:

\begin{subequations}
\begin{align}
\sum_{k: j \rightarrow k} P_{jk} &= P_{ij} + p_j && \forall j \in \mathcal{N} \\
\sum_{k: j \rightarrow k} Q_{jk} &= Q_{ij} + q_j && \forall j \in \mathcal{N} \\
v_j - v_k &= 2 \left( r_{jk} P_{jk} + x_{jk} Q_{jk} \right) && \forall (j \rightarrow k) \in \mathcal{E}
\end{align}
\end{subequations}

For the distribution grid $\mathcal{G}$ of interest, the adjacency matrix is a square $n\times n$ matrix $\tilde{M}$ such that its element $\tilde{M}_{ij}$ is 1 when there is an undirected edge from vertex $j$ to vertex $k$, and zero when there is no edge. An element in $\tilde{M}$ is defined as follows for an undirected graph:

\begin{eqnarray} 
\label{form1}
\tilde{M}_{ij}=
\begin{cases}
1& \text{if node $u_j$ is connected to node $u_k$};\\
0& \text{else}.\\
\end{cases}
\end{eqnarray}


Based on the topology of the networks, the distribution
networks are classified as either radial distribution networks or ring main distribution networks. The major difference lies on the fact that in the ring-based grid, there may be multiple paths between substations and loads. However, it is noted that during actual network operation, the tie-switches and circuit breakers are configured such that only one source feeds a load, and an electrically active path between them is unique. Hence, the active steady-state network can still be considered to be radial~\cite{pappu2017identifying}. The techniques developed herein are thus focused on radial networks, while it is possible to extend to ring structure with tailored branch.

\subsection{Power Measurements and Power Balance}
As described above, we now assume that the power measurements are taken from the same smart meter in the same area. The purpose is to eliminate the error caused by different regions and different types of smart electricity meters. We assume synchronized load data are obtained in kW for each node at multiple time steps.
We can organize the data of each time into a data matrix $X$:

\begin{eqnarray} 
\label{form2}
X(k) = [x_1(k) \; x_2(k)\; ... \; x_n(k)];
\end{eqnarray}
Where $X(k)$ is the data matrix of power measurements of $n$ nodes in the $k^{th}$ time interval; $x_i(k)$ is the power measured at $i^{th}$ node in the $k^{th}$ time interval. We note that in most common case, the active power measurements by smart meters are available, while some meters also have reactive power readings. We assume such measurements are complete for all nodes in the underlying grid for total of $K$ timesteps, but noise could exist for all measurements.

\begin{figure}[htbp]
\centering
\includegraphics[width=0.4\textwidth,height=0.35\textwidth]{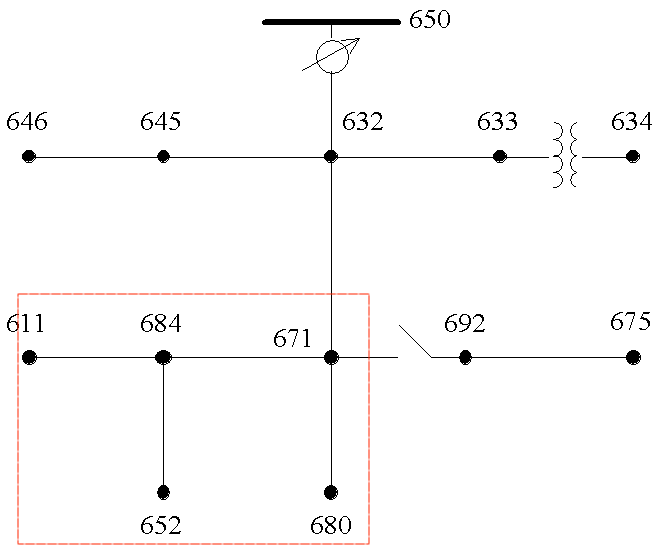}
\caption{13 Node Test Feeder data}
\label{fig1}
\end{figure}

We take the example of the IEEE-13 Node Test Feeder data (Figure \ref{fig1}) in the Radial Distribution Test Feeders \cite{kersting1991radial}.

In the dashed red box, consider we have a graph of a power network having five power measurements, connected through four power lines as shown in Figure \ref{fig2}. 
This allows us to calculate the power of the node pointed at by the arrow is equal to the sum of the power of all the nodes pointing at it. And without loss of generality, we assume each node has an independent power consumption. For instance $x_{684}$ denotes the node's individual power consumption, while $x'_{684}$ is the total power consumption for all the child nodes and the node itself. We can also treat the power summation $x_j'$ as the net injections to node $j$ via upstream power flow. The relationship of power measurements for the subnetwork shown in Figure \ref{fig1}(a) can be represented as follows:

\begin{figure}[htbp]
\centering
\includegraphics[width=0.45\textwidth]{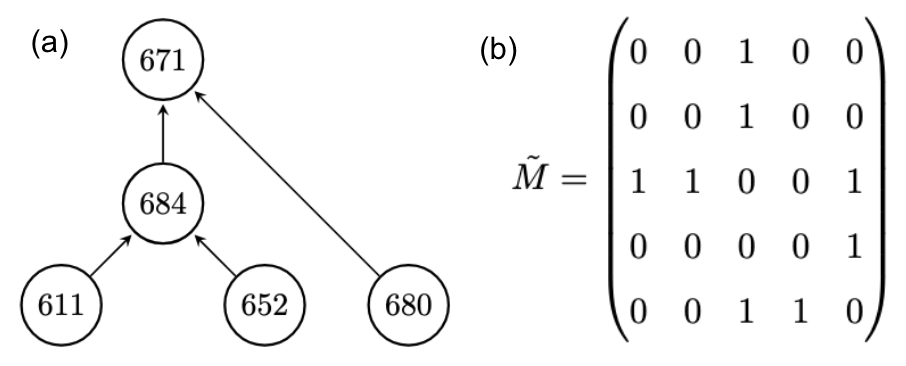}
\caption{(a). A toy example of distribution network with five nodes connected through four power lines. Each node is equipped with one power meter. (b). Corresponding adjacency matrix $\tilde{M}$.}
\label{fig2}
\end{figure}

\begin{subequations}
\begin{align}
\label{form3}
x_{684}^{\prime}(k)=x_{684}(k)+x_{611}(k)+x_{652}(k);\\
\label{form4}
x_{671}^{\prime}(k)=x_{684}^{\prime}(k)+x_{671}(k)+x_{680}(k).
\end{align}
\end{subequations}

We can then represent the graph  using an adjacency matrix in Fig \ref{fig2} (b), with node index in the sequence of $\{611,652,684,680,671\}$.

Knowing the child node and adjacency matrix, we can obtain the parent node's total power injections as:
\begin{eqnarray} 
\label{form8}
x'_j(k)=\tilde{M}_j x(k),
\end{eqnarray}
where $\tilde{M}_j$ denotes the $j$'s row of the adjacency matrix. 


\subsection{Distribution Network As a Tree}
\begin{figure}[htbp]
\centering
\includegraphics[width=0.45\textwidth]{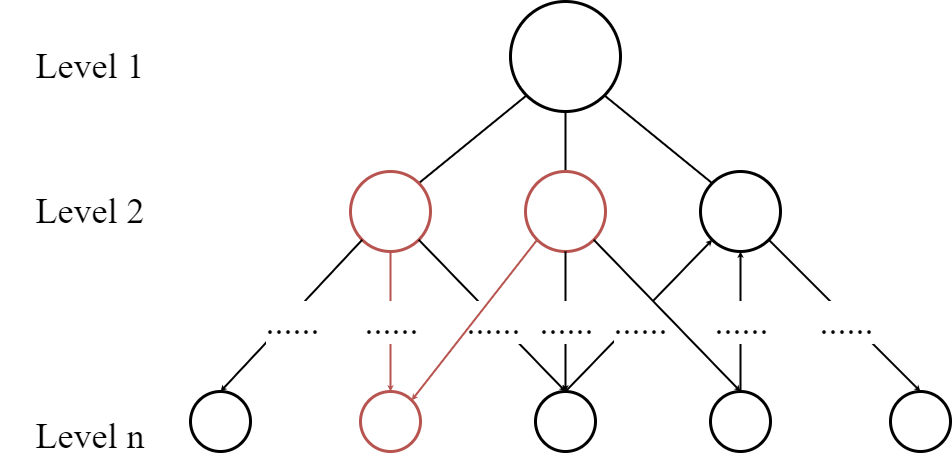}
\caption{Tree representation of network topology.}
\label{fig4}
\end{figure}

In graph theory, a tree is an undirected graph in which any two vertices are connected by exactly one path, or equivalently a connected acyclic undirected graph. A forest is an undirected graph in which any two vertices are connected by at most one path, or equivalently an acyclic undirected graph, or equivalently a disjoint union of trees \cite{deka2017structure, pappu2017identifying}. A tree network is composed of multiple layers of longitudinally connected star structures. Each node of the tree is either a computer or an adapter. While the topology of the real distribution network can be seen as a tree, connected by substations, feeders, transformers and instruments on the user power supply, the connections between them can be represented as edges, and the path from the substation to each user is unique and its structure is shown in Figure \ref{fig4}.

Substation nodes at Level 1 stands for the highest Level, while Level 2 to Level $n-1$ refers to feeders and transformers nodes at the middle Level, and consumer nodes at the bottom Level. In general, the closer the node devices are to the root of the tree, the better the topology inference performance is.

When only measurement data is available, false identification of connectivity will have a great influence on the whole network. For instance, in the red part of topology in figure \ref{fig4}, a child node could be identified as child node of multiple parent nodes (or a parent node corresponds to more child nodes than ground truth). In such cases, pure data-driven methods like PCA may cause a sequence of wrong connectivity identification. It has a great impact on the whole power grid structure, and this motivates us to achieve one-to-one correspondence between child nodes and parent nodes by fully utilizing the smart meter data and the tree structure in distribution grids.

\section{Fast Topology Identification with Subset Sum}

\subsection{Subset Sum}
The subset sum problem (SSP) is a classical combinatorial algorithm~\cite{kleinberg2006algorithm}. In its most general formulation, with a given set $X:=\{x_1, x_2,..., x_m\}$ consisting of $m$ integer elements, the task is to find the exact subset $S$, such that for the integer $G$, we can find a subset $S\subseteq X$, and the sum of integers in the subset is precisely $G$:
\begin{eqnarray} 
\label{form6}
\sum_{i\in S}{x_i} = G. 
\end{eqnarray}

There has been a vast literature in finding exact or approximate solutions of subset problem in a timely manner~\cite{antonopoulos2023faster}. And in this paper, we resort to a dynamic programming approach, as SSP can be solved with dynamic programming in pseudo-polynomial time. Dynamic programming is a method of solving complex problems by breaking the original problem into relatively simple subproblems. Dynamic Programming (DP) is often applied to problems with overlapping subproblems and optimal substructural properties.
In this paper, we generalize the standard algorithm to include non-integer energy readings, and find the case where the sum of the first $|R|$ elements in a set $X$ is equal to $G$. And this will recover the non-zero entries in the adjacency matrix $\tilde{M}$. The complexity of DP-based subset sum is of $O(n\log n+ nK|S|)$, where first term denotes the sorting time, and $|S|$ denotes the number of parent nodes used for calculating the summation.

At the same time, we improve the shortcoming that Dynamic Programming stops when it finds a set of answers that meet the criteria. The proposed algorithm can find all possible combinations of tree structures that meet the criteria in the subset. As there are multiple group of measurements available, we also include a majority vote step at the end of the process, so that the tree structure with the best fit is voted as the underlying topology.

\subsection{Losses and Measurement Errors}
In the latest ANSI-2015 standard, it states that electricity meters must be accurate to within +/- 0.1 $\%$, +/- 0.2$\%$, and +/- 0.5$\%$ of the true value at full load, of true values for 0.1, 0.2 and 0.5 accuracy class meters, respectively \cite{ANSI2015}.

We model this error as a Gaussian distribution, where the error vector due to random errors in smart meters is distributed as $\epsilon{(k)}$ in the $k^{th}$ time interval.
The size of $\epsilon{(k)}$ is $1\times n$, where each element of $\epsilon_i{(k)}$ follows the Gaussian distribution of $\epsilon_i{(k)} \sim (0,\sigma)$, and $\sigma$ is the variance of errors taking one of the three values: 0.1$\%$, 0.2$\%$ or 0.5$\%$.

\subsection{Topology Identification through HSSP}
The basis of topology identification for the tree structure of power grid is to implement the subset sum algorithm in every two directly connected levels. In practice, nodes in the distribution network can be treated as residing in different layers of subtrees. And each layer operates at a known voltage level. Motivated by this observation,  we redesign subset sum algorithm to accurately reveal such hierarchical tree structures.

Any set of two consecutive layers, when visualized separately, appears as a forest of oriented trees. In this case, the problem of identifying the multi-layer topology can be reduced to finding the connectivity of the directed tree forest. By inferring the connectivity between all possible successive layers, the complete network topology can be identified. 

The standard subset sum algorithm can only find the relationship between child nodes and the direct parents. In this paper, we improve the subset sum algorithm to a hierarchical subset sum algorithm (HSSP), so that it can recognize the multi-layer topology. Essentially, we keep an identifier for the level of each node $j$, denoted as $l_j$. Once a new relation between two nodes $j$ and $k$ are discovered,  nodes can be interpreted as either node or parent in the tree depending on $l_j$ and $l_k$. Such identifier helps reduce search space, as two nodes which are identified impossible to be parent and son nodes will be excluded from next round of subset sum calculation. This will further reduce the topology identification errors, while it can also integrate structural information if they are available. 

The implementation algorithm is shown in Algorithm \ref{alg:1}. We note that the hierarchy information in Line 1 of the algorithm is optional, and the algorithm can be applied without restrictions on network size and number of data samples.

\begin{algorithm}[h]
	\caption{ Topology Identification with HSSP}
	\label{alg:1}
	\begin{algorithmic}[1]
	    \REQUIRE Timesteps $K$, distribution grid with $n$ nodes~~\\
Data measurements of size $n\times K$;\\
		 Initialize $X$ for all data entries;\\
   $\tilde{m}_{n*n}=\bm{0}$;
		\STATE Divide $X$ into $X_i$ and $X_d$ based on grid structure of $X$;
		\STATE $j=0$; $j$ is the index in $X_d$;
		\FOR{$x_j$ $\in$ $X_d$}
		\STATE For each node in $x_d$, for $k\in \{0,...,K-1\}$, find one and more nodes in $x_i(k)$ whose sum is equal to $x_d(k)$; 
		\STATE Vote for the subset that appears the most and get the possible subsets $S$;
		\STATE $j=j+1$; 
		\ENDFOR
		\IF{the independent node $(G_i)$ belongs to parent node $(G_d)$} 
		\STATE $\tilde{m}_{id} = \tilde{m}_{di} = 1$;
		\ENDIF 
  \STATE Reduce redundant parents nodes by votes and hierarchy;
		\ENSURE ~~\\ 
		Compare  $\tilde{m}$ to $\tilde{M}$; Calculate identification accuracy;\\
	   \textbf{End}
	\end{algorithmic}  
\end{algorithm}

\section{Simulation Results}
\subsection{Plain Subset Sum Algorithm}
In this Section, we demonstrate how proposed method can be implemented to achieve fast and accurate topology identification using few samples of smart meter data. We also open-source our method for future applications and benchmarking~\footnote{\url{https://github.com/michellexu99/HSSP}}. 

Since the noiseless case is trivial, we generated noisy data for different networks and the results are presented. In the first simulation example, the network was built randomly using the random number generators in Python.
First, we compared the accuracy and running time of PCA algorithm and HSSP algorithm in the case of no hierarchy, the detailed settings are as follows:

\begin{itemize}
    \item Randomly generating data for 20 time periods, each time period containing 10 child nodes, and parent nodes power measurements are obtained by multiplying the child node with the adjacency matrix.
    \item Value range of all nodes are chosen randomly (uniformly) between 0 and 50.
    \item We add a table reading error which follows a Gaussian distribution with variance 0 and mean 0.01.
\end{itemize}

In the case of using simulated data, we found that the accuracy of both algorithms is 1.0, but the operation time of HSSP algorithm is 1.285 second on average, nearly 0.8 seconds less than that of PCA algorithm. This indicates that HSSP algorithm is faster and requires less time than PCA.

\subsection{Subset Sum on Hierarchical Tree}
However, the above topology identification results use non-hierarchical data, and most of the distribution grid in real life is still having a hierarchical tree structure. The simulation setup for a set of IEEE distribution grid test feeders is as follows:
\begin{itemize}
    \item Randomly generating data for 10 time periods, each time period contains $N$ power measurements for $N$ nodes, where $N$ is equal to $13(N1), 33(N2), 63(N3), 93(N4) $ and $ 123(N5)$. Parent node readings are obtained by multiplying the child node with the adjacency matrix.
    \item Smart meter readings of all nodes are chosen randomly (uniformly) between 25 and 50.
    \item We inject noise $\epsilon$ to meter readings, and such error follows a Gaussian distribution with zero mean, and variance ranging from 0.01, 0.02, 0.05 to 2. The value of 2 is used to verify the proposed method's performance under extreme cases.
\end{itemize}

\begin{table}[h]
\centering
\caption{Comparison of the accuracy of PCA and HSSP algorithms in the case of hierarchy}
\label{table-three}
\begin{minipage}[c]{0.25\textwidth}
\resizebox{1\textwidth}{!}{
\begin{tabular}{c|c|c|c}
\hline{} Size &\textbf{$\epsilon$} &\textbf{\makecell[c]{PCA \\based method}} &\textbf{\makecell[c]{HSSP \\based method}} \\ 
\hline 
\multirow{4}*{$N1$} & 0.01 & 0.2971 & 1.0  \\
\cline{2-4}& 0.02 & 0.0595 & 1.0 \\
\cline{2-4}& 0.05 & 0.2738 & 1.0 \\
\cline{2-4}& 2.0 & 0.1667 & 1.0 \\
\hline
\multirow{4}*{$N2$} & 0.01 & 0.1255 & 0.9761  \\
\cline{2-4}& 0.02 & 0.5672 & 0.9376 \\
\cline{2-4}& 0.05 & 0.3763 & 0.9743 \\
\cline{2-4}& 2.0 & 0.3369 & 0.9339 \\
\hline
\multirow{4}*{$N3$} & 0.01 & 0.2031 & 0.9567  \\
\cline{2-4}& 0.02 & 0.3180 & 0.9376 \\
\cline{2-4}& 0.05 & 0.1412 & 0.9743 \\
\cline{2-4}& 2.0 & 0.3956 & 0.9461 \\
\hline
\end{tabular}}
\end{minipage}
\begin{minipage}[c]{0.22\textwidth}
\resizebox{1.1\textwidth}{!}{
\begin{tabular}{c|c|c|c}
\hline{}Size &\textbf{$\epsilon$} &\textbf{\makecell[c]{PCA \\based method}} &\textbf{\makecell[c]{HSSP \\based method}} \\ 
\hline 
\multirow{4}*{$N4$} & 0.01 & 0.4697 & 0.9694  \\
\cline{2-4}& 0.02 & 0.2894 & 0.9673 \\
\cline{2-4}& 0.05 & 0.2903 & 0.9656 \\
\cline{2-4}& 2.0 & 0.4403 & 0.9704 \\
\hline
\multirow{4}*{$N5$} & 0.01 & 0.3559 & 0.9735  \\
\cline{2-4}& 0.02 & 0.3788 & 0.9674 \\
\cline{2-4}& 0.05 & 0.2337 & 0.9698 \\
\cline{2-4}& 2.0 & 0.3411 & 0.9764 \\
\hline
\end{tabular}}
\end{minipage}
\end{table}
As can be seen from Table \ref{table-three}, the performance of PCA varies a lot under different settings. PCA is also not robust to measurement noises, and recover entirely wrong graph structure when $\epsilon\geq 0.05$. On the other side, measurement error $\epsilon$ has slight impact on the accuracy of the HSSP algorithm. With the increase of the measurement error value, the accuracy of the HSSP algorithm generally shows a downward trend, but the decline is very mild, and the accuracy is still high when encountering extreme measurement noises $\epsilon=2$, indicating the robustness of the proposed approach. 
It can be seen that the accuracy of using HSSP algorithm to identify multi-level grid topology decreases with the increase of $N$, but the overall result is much better compared to PCA.

Next we evaluate how the hierarchial order help topology recovery empirically. To achieve so, we disorder the nodes and remove the identifier for each tree layer, and compare the accuracy of HSSP algorithm for topology identification under ordered and unordered scenarios.

\begin{figure}[!htbp]
\centering
\begin{tikzpicture}[scale=0.65]
\begin{axis}
[ybar,
    xlabel={$N$},
    ylabel={$Accuracy$  $\%$},
    symbolic x coords={13, 33, 63, 93, 123},  
            xtick=data ]
\addplot[draw=blue,fill=cyan] 
coordinates
{(13,1.0) (33,0.9761)
 (63,0.9582) (93,0.9694) (123,0.9735)
};
\addplot[draw=black,fill=blue] 
coordinates
{(13,0.8462) (33,0.9027)
 (63,0.9567) (93,0.9423) (123,0.9635)
};
\legend{Hierarchy, No hierarchy
}
\end{axis}
\end{tikzpicture}
\caption{Accuracy of HSSP algorithm in topology identification under ordered and unordered nodes.\vspace{-20pt} }
\label{fig6}
\end{figure}
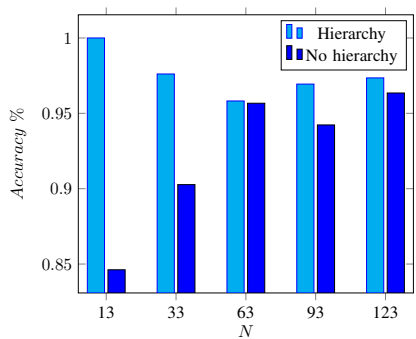

We test if the hierarchical information can impact the performance of subset sum algorithm. As can be seen from Figure \ref{fig6}, the accuracy of non-hierarchical data is slightly lower than that of hierarchical data, but with the increase of $N$, the accuracy of the two will be closer. This may be due to the fact that increasing dimension of data reduces the number of similar subsets, making the proposed approach easier to make the right topology estimation decisions.

\vspace{-10pt}
\section{Conclusion}
In this paper, we propose a novel hierarchical subset sum algorithm for identifying the connectivity of low-voltage distribution networks. This method  is inspired by the standard subset sum algorithm, and can be generalized to fast multi-level tree network topology recovery. By using standard distribution feeders and testing on noisy measurements, we find that the HSSP algorithm performs well in identifying hierarchical structure. In the future work, we will consider additional information, such as phase angles provided by phasor measurement units (PMUs), and solve the topology identification problem with partially observable measurements. Larger graph, ring or loop connections, and interconnection with transmission grid will be also taken into account.

\bibliographystyle{IEEEtran}
\bibliography{bib}

\begin{thebibliography}{10}
\providecommand{\url}[1]{#1}
\csname url@samestyle\endcsname
\providecommand{\newblock}{\relax}
\providecommand{\bibinfo}[2]{#2}
\providecommand{\BIBentrySTDinterwordspacing}{\spaceskip=0pt\relax}
\providecommand{\BIBentryALTinterwordstretchfactor}{4}
\providecommand{\BIBentryALTinterwordspacing}{\spaceskip=\fontdimen2\font plus
\BIBentryALTinterwordstretchfactor\fontdimen3\font minus
  \fontdimen4\font\relax}
\providecommand{\BIBforeignlanguage}[2]{{%
\expandafter\ifx\csname l@#1\endcsname\relax
\typeout{** WARNING: IEEEtran.bst: No hyphenation pattern has been}%
\typeout{** loaded for the language `#1'. Using the pattern for}%
\typeout{** the default language instead.}%
\else
\language=\csname l@#1\endcsname
\fi
#2}}
\providecommand{\BIBdecl}{\relax}
\BIBdecl

\bibitem{william2002distribution}
H.~K. William \emph{et~al.}, ``Distribution system modeling and analysis,''
  \emph{Ed.: CRC Press, USA}, 2002.

\bibitem{fan2009evolution}
J.~Fan and S.~Borlase, ``The evolution of distribution,'' \emph{IEEE Power and
  Energy magazine}, vol.~7, no.~2, pp. 63--68, 2009.

\bibitem{deka2017structure}
D.~Deka, S.~Backhaus, and M.~Chertkov, ``Structure learning in power
  distribution networks,'' \emph{IEEE Transactions on Control of Network
  Systems}, vol.~5, no.~3, pp. 1061--1074, 2017.

\bibitem{lueken2012distribution}
C.~Lueken, P.~M. Carvalho, and J.~Apt, ``Distribution grid reconfiguration
  reduces power losses and helps integrate renewables,'' \emph{Energy Policy},
  vol.~48, pp. 260--273, 2012.

\bibitem{chembe2009reduction}
D.~K. Chembe, ``Reduction of power losses using phase load balancing method in
  power networks,'' in \emph{Proceedings of the World Congress on Engineering
  and Computer Science}, vol.~1, 2009, pp. 20--22.

\bibitem{jayadev2016novel}
S.~P. Jayadev, A.~Rajeswaran, N.~P. Bhatt, and R.~Pasumarthy, ``A novel
  approach for phase identification in smart grids using graph theory and
  principal component analysis,'' in \emph{2016 American Control Conference
  (ACC)}.\hskip 1em plus 0.5em minus 0.4em\relax IEEE, 2016, pp. 5026--5031.

\bibitem{mabrouk2022distribution}
A.~Mabrouk and R.~Rajagopal, ``Distribution grid topology estimation: A new
  approach-based on bayesian network models,'' in \emph{2022 IEEE Eighth
  International Conference on Big Data Computing Service and Applications
  (BigDataService)}.\hskip 1em plus 0.5em minus 0.4em\relax IEEE, 2022, pp.
  124--131.

\bibitem{li2013blind}
X.~Li, H.~V. Poor, and A.~Scaglione, ``Blind topology identification for power
  systems,'' in \emph{2013 IEEE International Conference on Smart Grid
  Communications (SmartGridComm)}.\hskip 1em plus 0.5em minus 0.4em\relax IEEE,
  2013, pp. 91--96.

\bibitem{ardakanian2019identification}
O.~Ardakanian, V.~W. Wong, R.~Dobbe, S.~H. Low, A.~von Meier, C.~J. Tomlin, and
  Y.~Yuan, ``On identification of distribution grids,'' \emph{IEEE Transactions
  on Control of Network Systems}, vol.~6, no.~3, pp. 950--960, 2019.

\bibitem{karimi2021joint}
H.~S. Karimi and B.~Natarajan, ``Joint topology identification and state
  estimation in unobservable distribution grids,'' \emph{IEEE Transactions on
  Smart Grid}, vol.~12, no.~6, pp. 5299--5309, 2021.

\bibitem{cavraro2018graph}
G.~Cavraro and V.~Kekatos, ``Graph algorithms for topology identification using
  power grid probing,'' \emph{IEEE control systems letters}, vol.~2, no.~4, pp.
  689--694, 2018.

\bibitem{pappu2017identifying}
S.~J. Pappu, N.~Bhatt, R.~Pasumarthy, and A.~Rajeswaran, ``Identifying topology
  of low voltage distribution networks based on smart meter data,'' \emph{IEEE
  Transactions on Smart Grid}, vol.~9, no.~5, pp. 5113--5122, 2017.

\bibitem{baran1989optimal}
M.~Baran and F.~F. Wu, ``Optimal sizing of capacitors placed on a radial
  distribution system,'' \emph{IEEE Transactions on power Delivery}, vol.~4,
  no.~1, pp. 735--743, 1989.

\bibitem{kersting1991radial}
W.~H. Kersting, ``Radial distribution test feeders,'' \emph{IEEE Transactions
  on Power Systems}, vol.~6, no.~3, pp. 975--985, 1991.

\bibitem{kleinberg2006algorithm}
J.~Kleinberg and E.~Tardos, \emph{Algorithm design}.\hskip 1em plus 0.5em minus
  0.4em\relax Pearson Education India, 2006.

\bibitem{antonopoulos2023faster}
A.~Antonopoulos, A.~Pagourtzis, S.~Petsalakis, and M.~Vasilakis, ``Faster
  algorithms for k-subset sum and variations,'' \emph{Journal of Combinatorial
  Optimization}, vol.~45, no.~1, p.~24, 2023.

\bibitem{ANSI2015}
N.~E. M.~A. [NEMA], ``Electric meters - code for electricity metering
  (incorporates ansi c12.20-2015),''
  \url{https://webstore.ansi.org/Standards/NEMA/ANSIC122022}.

\end{thebibliography}

\end{document}